\documentclass{PoS}

\input{psfig}
\usepackage[dvips]{epsfig}

\def\kms    {\ifmmode {\rm km\,s}^{-1} \else km\,s$^{-1}$\fi}
\def\mujybm {${\rm \mu}$Jy\,beam$^{-1}$}

\def\degr   {\hbox{$^\circ$}}

\title{Wide-field VLBI imaging of M31 - first results}

\ShortTitle{Wide-field VLBI imaging of M31 - first results}

\author{\speaker{Megan Argo}\\
        ICRAR / Curtin University, GPO Box U1987, Perth, WA 6845, Australia\\
        E-mail: \email{m.argo@curtin.edu.au}}

\author{John Morgan\\
        ICRAR / Curtin University, GPO Box U1987, Perth, WA 6845, Australia\\
        E-mail: \email{j.morgan@curtin.edu.au}}

\author{Emil Lenc\\
        CSIRO Astronomy and Space Science, PO Box 76, Epping NSW 1710, Australia\\
        E-mail: \email{Emil.Lenc@csiro.au}}

\author{Olaf Wucknitz\\
	Argelander-Institut f\"ur Astronomie, Auf dem H\"ugel 71, D-53121 Bonn, Germany\\
	E-mail: \email{wucknitz@astro.uni-bonn.de}}

\author{Franco Mantovani\\
	Istituto di Radioastronomia - INAF, via Gobetti 101, 40129 Bologna, Italy\\
	E-mail: \email{f.mantovani@ira.inaf.it}}

\author{Danielle Fenech\\
	Department of Physics and Astronomy, University College London, London\\
	E-mail: \email{dmf@star.ucl.ac.uk}}

\author{Steven Tingay\\
        ICRAR / Curtin University, GPO Box U1987, Perth, WA 6845, Australia\\
        E-mail: \email{steven.tingay@icrar.org}}

\abstract{One of our closest neighbours, the Andromeda Galaxy (M31) has been the subject of numerous large area studies across the entire spectrum, but so far full-disk radio surveys have been conducted only at low resolution.  The new wide-field capabilities of the DiFX software correlator present the possibility of imaging the entire primary beam of a VLBI array, thus enabling a high-resolution wide-field study of the entire galaxy.  Using the VLBA and EVN, pilot observations of M31 have been carried out with the aim of using these new wide-field techniques to characterise the population of compact components at VLBI resolution both within and behind one of our nearest neighbours.  This contribution describes the observations carried out, the preliminary processing and first results.
}

\FullConference{10th European VLBI Network Symposium and EVN Users Meeting: VLBI and the new generation of radio arrays \\
		 September 20-24, 2010\\
		 Manchester, UK}

\begin{document}

\section{Wide-field imaging}

Traditionally, VLBI is not well-suited to wide-field imaging due to time and bandwidth smearing effects which become worse with greater averaging.  Making larger images involves correlating with narrow channels and short integrations which can result in very large datasets and is often limited by the output rate of a given correlator.  A single UV dataset with sufficient time and spectral resolution to avoid smearing across the entire primary beam of the VLBA would be extremely large and unmanageable in AIPS.

Software correlators allow for much more flexibility, limited mainly by the available computing power.  The DiFX software correlator \cite{deller07} is one such correlator, now in use as the production correlator for both the Australian LBA and the VLBA.  The new wide-field capabilities of DiFX (correlating a large number of phase centres in a single correlation pass) allow the end user to produce far more manageable output datasets and presents the possibility of imaging the entire primary beam of a VLBI array, greatly increasing the information obtainable from a single observation and speeding up surveys.  For details of the new capabilities of DiFX, see \cite{morgan10a, deller10}.

\section{Why M31?}

M31 covers a large area on the sky, its disk has a major axis of some 4\degr\ in the radio.  It has a bright core of emission, a 10\,kpc ring of star formation which shows up strongly not only in the infra-red and ultra-violet, but also in low resolution radio surveys.  There is a history of star formation within the spiral arms with regular optical novae detections, the occasional supernova, and many known H{\sc ii} regions, X-ray sources, and numerous background AGN candidates.

One of the closest galaxies to the Milky Way, M31 has been surveyed extensively across all wavebands, including several low-resolution radio studies carried out over the last 50 years.  Braun \cite{braun90} surveyed the eastern half of the disk with the VLA in B-, C- and D-configurations, resulting in a detailed map at a resolution of 5" and a source list containing 534 objects down to a limit of 35\,\mujybm at L-band.  Beck et al \cite{beck98} surveyed the disk of the galaxy at L-band using the VLA in D-configuration, covering the entire 10\,kpc ring in seven pointings at a resolution of 45", resulting in an additional catalogue of 36 linearly polarised background sources to a limit of 75\,\mujybm.
Of the catalogued radio sources, a large fraction are unresolved on VLA scales.  Of the 534 continuum sources in Braun's survey \cite{braun90}, 103 have possible associations with H$\alpha$ nebulae and seven are co-located with previously known supernova remnants.  The map also contains 121 objects associated with previously known radio sources at 408\,MHz, 610\,MHz or 1.4\,GHz, many of which appear to be background AGN, some with classic core-jet or double lobed structures.

The star formation rate in M31 has been measured using various methods and has been estimated to be around 0.5 to 1 M$_{\odot}\,{\rm yr}^{-1}$, significantly less than in nearby starbursts.  Estimates of the supernova rate are varied and have significant uncertainty.  At the distance of M31, VLBA observations at 1.4\,GHz give a linear resolution of $\sim$0.02\,pc which would allow even young supernova remnants to be resolved.

The scientific aim of this project is to produce a catalogue of compact components within the M31 field, both those within and behind the disk, in order to investigate both star formation within the galaxy and the population of background AGN.  The observations described here are essentially a pilot study in preparation for imaging the entire disk of the galaxy at VLBI resolution, and are also testing wide-field techniques which will become commonplace for future instruments.

\section{Observations}

A field in the eastern half of M31 was observed at 1.4\,GHz with the European VLBI Network on June 7$^{\rm th}$ 2010.  Due to the heterogeneous nature of the array, this observation used four pointings spaced 7.5' apart on the sky, Nyquist sampling the sky with the Effelsberg beam in order to increase the size of the central sensitive area at an acceptable cost in overall sensitivity.  The run was 8 hours in length at a data rate of 512\,Mbps.  These data have not yet been processed.  For further discussion of multiple pointing VLBI surveys, see \cite{morgan10b}.

\begin{figure*}
\centering
\includegraphics[width=11cm]{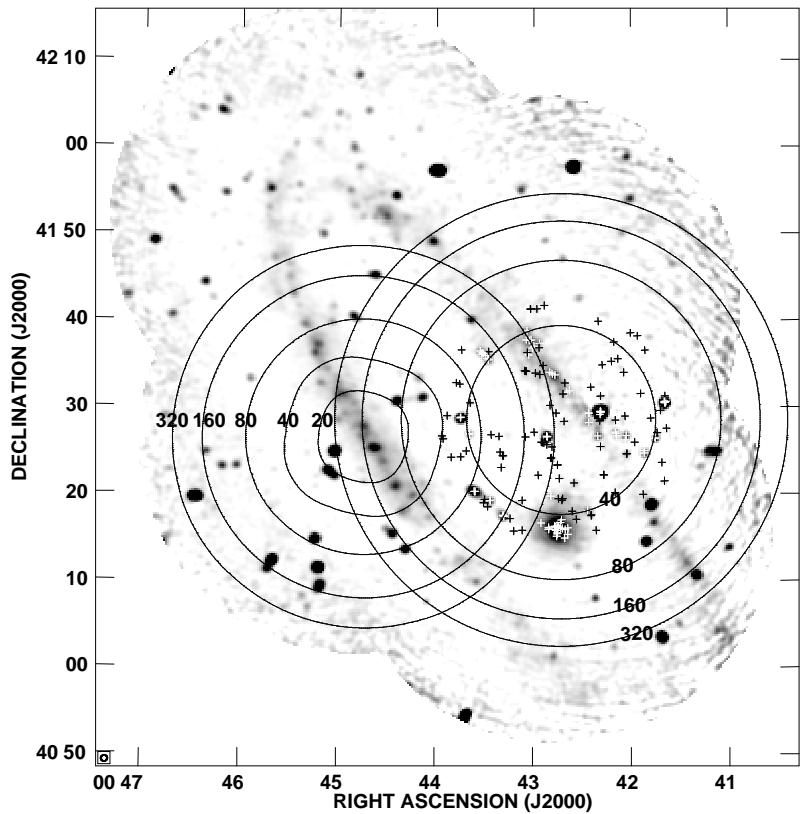}
\caption[M31 observed fields]
{\label{fig_fields}The EVN (left) and VLBA (right) fields observed as part of this study.  The greyscale plot is the 1.4-GHz VLA data from Beck et al \cite{beck98}, contours are the predicted sensitivities for the VLBI observations in \mujybm\ \cite{morgan10b} and crosses show the phase centres used in the first correlation pass with the VLBA data.}
\end{figure*}

An adjacent field was also observed using a single pointing of the Very Long Baseline Array.  This observation utilised all ten stations of the array, had a total length of 8 hours, used 4 IFs of 16 MHz each, nodding to a phase calibrator, J0038+4137, located roughly 1\degr\ away, and was observed on July 4$^{\rm th}$ 2010, again at 1.4\,GHz.  Both fields are illustrated in Figure \ref{fig_fields}.  The data from this observation were correlated with DiFX in Socorro as a regular continuum experiment, but the baseband data were also copied to disk and shipped to Curtin University where they were loaded onto a cluster for re-correlation and processing.

\section{Targets}

The aim is to map the entire primary beam, but for an initial test run with the VLBA data only a limited number of fields were used so that procedures could be tested quickly with a small amount of data.

In source catalogues from NED\footnote{NASA's Extragalactic Database, http://nedwww.ipac.caltech.edu/} there are many hundreds of sources listed in the literature within 0.5\degr\ of the VLBA pointing centre.  Applying various cuts and excluding duplicates and those sources not known to be radio bright resulted in a list of 175 sources to use as a first pass.  Figure \ref{fig_fields} also shows the location of the target sources superimposed on a low resolution map of M31.  Our first run used just these positions as phase centres, most of which are known radio sources from the Braun and Beck et al surveys \cite{braun90,beck98}.

\section{Processing}

The baseband data from the VLBA observation were shipped to Curtin University on disk, as well as being correlated in Socorro as usual using standard procedures.
The maturity of the DiFX correlator, along with the fact that the data had already been correlated at the VLBA (meaning that all parameters required for correlation, such as clock offsets, had been determined) essentially turned re-correlation into a high performance computing problem.  Re-correlation was carried out on a cluster at Curtin University using the new multiple-phase centre capabilities of DiFX, producing a single manageable $uv$-fits file per target phase centre.

Calibration of the re-correlated datasets was then straightforward and proceeded as for a standard VLBA continuum observation.  The calibrator scans from Socorro were used to produce calibration tables which could then be applied to each re-correlated dataset in the same way calibration is transferred from an averaged dataset in a spectral line experiment.  The re-correlated data had a greater spectral resolution than the data from the Socorro correlator (with 128 channels per IF, rather than 32), so the bandpass calibrator data also had to be re-correlated at Curtin with the same parameters as the target scans in order to perform a bandpass calibration.

Flagging of bad times and antennas was also carried out on the data correlated in Socorro and transferred, but bad channels were flagged on the re-correlated data.  In particular, the first two IFs suffered significantly from RFI.  While most calibration is equally applicable to all fields, it was found that when a flag table created for one field was transferred to a different phase centre, not all of the bad data were removed.  Adding flags for a number of fields well-spaced around the primary beam should improve this.

\section{Results}

\begin{figure*}
\centering
\includegraphics[width=7cm]{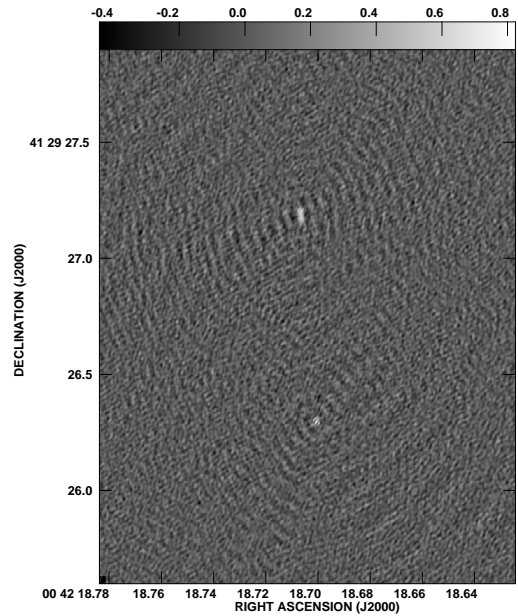}
\caption[M31 double source]
{\label{fig_double}The double source 4218+2926 detected in the VLBA observations, located north west of the nucleus, just outside the 10\,kpc ring.}
\end{figure*}

Figure \ref{fig_fields} shows the positions of each of the phase centres re-correlated at Curtin.  The most efficient source-searching technique is to pipeline the production of dirty maps and output statistics on the brightest and rms pixel values in each field.  This output suggested a couple of possible detections which were then examined by eye.  Each pipelined image was also examined by eye and, while most of the images were empty, a total of three sources were detected at $\geq5\sigma$:

\begin{itemize}

\item {\bf 4145+2623} - a 5.2\,mJy source in the NVSS \cite{condon98}, detected here at 1\,mJy (5$\sigma$), located outside the 10\,kpc ring to the west;

\item {\bf 4251+2633} - a background AGN candidate with a flux of 24\,mJy in the NVSS and 22\,mJy in \cite{beck98}, detected in this observation with a flux of 2\,mJy, located to the north of the nucleus and inside the ring on the sky;

\item {\bf 4218+2926} - located just outside the ring, another background AGN candidate and one of the brightest radio sources within the whole VLBA field with a flux of 372\,mJy in the NVSS and 307\,mJy in \cite{beck98}, detected here as a double source with the components having fluxes of 1 and 2\,mJy respectively, see Figure \ref{fig_double}.

\end{itemize}

Much of the emission seen in VLA or single-dish maps is extended, so not all of the sources detected in previous surveys would necessarily be detected at this angular resolution.  Given the number of AGN candidates and other known radio sources which are compact on VLA scales, however, the detection rate is lower than might have been expected.

One factor is that the list of phase centres actually included many overlapping fields, so the total number of independent fields imaged is considerably less than 175; the three sources detected were each present in several images due to overlap.  Another factor is the noise level.  Although the predicted noise level near the centre of the primary beam is $\sim$40\,\mujybm, the actual noise level in the pipelined images was $\sim$0.2\,mJy\,beam$^{-1}$.  This can be improved by refining the calibration.  There could also be other, astrophysical, effects which reduce the detection rate, such as angular broadening due to the emission from these sources passing through the ISM of an entire (and reasonably inclined) galaxy.

It was hoped that there would be at least one field containing a detection strong enough to be used as an in-beam calibrator which could then be used to refine the calibration further, but the sources detected so far are too weak to be used for this purpose.

\section{What's next?}

Initial results show that several sources have been detected in the M31 field in a single 8-hour VLBA observation, separated by up to 15' on the sky.  There are a number of ways in which the calibration can be refined which should improve the noise level in the pipelined images.  Once the pipeline is functioning satisfactorily, the entire primary beam can be imaged and searched for sources.

The procedures outlined here will also be applied to the EVN data.  This will be significantly more complicated, however.  Firstly, the EVN is a heterogeneous array, each antenna is different and will have a different response which must be taken into account in the calibration.  Secondly, in order to increase the sensitive area in the centre of the field, the observation used multiple pointings, Nyquist sampling the sky with the Effelsberg beam.

The work presented here illustrates that imaging anywhere within the primary beam of a VLBI array is not only possible, but given suitable computing facilities can be fairly straightforward for simple observations.

\acknowledgments
We would like to thank Walter Brisken and Adam Dellar (NRAO) for useful discussions and providing the baseband data, and Graham Jenkins (VPAC) for transferring the data.

\end{document}